# Low intensity noise high-power tunable fiber-based laser around 1007 nm

Benoît Gouhier[1], Clément Dixneuf [1,2], Adèle Hilico[1], Germain Guiraud[2], Nicholas Traynor[2] and Giorgio Santarelli[1]

[1]Laboratoire Photonique, Numérique et Nanosciences, Institut d'Optique Graduate School, CNRS, Université de Bordeaux, 1, Rue F. Mitterand, 33400 Talence, France

[2]Azur Light Systems, Avenue de la Canteranne, 33600 Pessac, France

*Abstract*— We report the development of a tunable, low-noise, all-fiber, single-frequency, linearly polarized high-power master oscillator power amplifier operating over 20 nm, from 1017 nm down to 997 nm. Using Ytterbium highly-doped silica fiber with large mode area, more than 3 W of 997 nm signal was obtained with less than 2 % of ASE. These figures keep improving with operating wavelength, to reach more than 16 W and 0.03 % of ASE at 1017 nm. The optical signal to noise ratio exceeds 40dB for the worst case up to more than 60dB. In addition, relative intensity noise remains very low over the whole tuning range (-130dBc@10kHz) down to -160dBc at 10 MHz. This robust laser has a strong potential to be a new workhorse for demanding applications in this spectral range.

Key words—laser, MOPA, fiber, single frequency

## 1. Introduction

Driven by relentless research efforts focused on overcoming technological hurdles, fiber lasers have established themselves as a very convincing choice for numerous applications requiring high-power, low-noise laser radiation. Because of their high modal quality, exemplary polarization properties, excellent power scaling and wide spectral operation range, Ytterbium (Yb)-doped fiber lasers are one of the workhorses of many scientific and industrial applications around 1 μm[1]–[3]. Using a master oscillator power amplifier (MOPA) architecture, it is also possible to achieve single-frequency (SF) operation at high power[4]–[8].

Although the emission spectrum of Yb ranges from 970 to 1200 nm[9]–[11], most developments – including SF lasers – happen either in the maximum gain region (1020-1080 nm)[8], [12], or at wavelengths with specific usage (e.g. 1178 nm for guide star[13], 1014.8 nm for mercury cooling[14, p. 7], [15, p. 8], …) where the investment overhead caused by the exotic wavelength is guaranteed to be paid for. Even the challenging three-level transition of the Yb ion at 976nm has been extensively explored, leading to mature industrial level lasers able to deliver in excess of 10 W [16]. However, some less popular spectral areas see little work, while they could benefit from the maturity of Yb fiber lasers. In particular, there is no report of a cost-effective, industrial-grade fiber laser around 1000 nm. The few studies in the spectral are focused on cryogenic systems[17] or rod-type fiber[18], lacking the robustness of monolithic fiber lasers. The best effort in this area is to credit to Yan Feng group which have realized an all-fiber MOPA system tunable down to 1008 nm, although with strong ASE content[19]. Solid-state (Ti:sapph), Yb[20]–[23]) or semiconductor solutions (VECSEL[24], [25], tapered amplifier[26]) also exist around these wavelengths; however, they remain either complex systems, or exhibit limited power (<2 W) or mediocre and asymmetric modal quality ($M^2$>1.5). Efficient lasing at short wavelengths i.e.< 1020nm has been obtained for multi-mode laser configurations with phosphate glass fibers. However, these remain difficult to splice and not as mature as silica-based fibers [27]. We have already demonstrated an optimized 50 W version of this laser at the wavelength of 1013nm [28]. But demanding applications such as quantum simulation[29], optical cooling[30], or stabilization on hyperfine iodine transitions at 501.7 nm[20] would clearly benefit from the development of a robust laser delivering spectrally pure, low-noise, SF radiation with a broader tuning range in a spectral region with very few high power sources.

## 2. Tunable laser description

In this paper, we detail the realization of a tunable, low-noise, all-fiber, single-frequency, linearly polarized high-power MOPA operating over 20 nm, from 1017 nm down to 997 nm. The output signal shows excellent properties, with very weak ASE and low intensity noise.

Because of its robust, all-fiber architecture and its ambient temperature operation, this type of laser has a very strong disruption potential for demanding applications in this spectral region.

The schematic of the system is shown in Fig. 1. A tunable SF source (Toptica DL Pro) delivering 80 mW of isolated, fiber-coupled light is spliced to an additional isolator for safety. The SF operation of the tunable laser has been confirmed by fast heterodyne beating with a narrower SF laser diode at 1025 nm. The linewidth of the Toptica laser was then estimated around 30 kHz, with very little change with emission wavelength. The light is then injected via a WDM into a 20 cm-long core-pumped pre-amplifier. The active fiber (nLight Liekki Yb 1200 6/125) is pumped with a 976 nm single-mode diode (Gooch and Housego) able to deliver up to 550 mW. Depending on the wavelength, the output power varies between 200 mW (at 997 nm) up to 350 mW (at 1017 nm). The lower spectral bound is set by the input WDM which exhibits more than 3 dB losses below 997 nm. A notch filter removing the ASE at the output sets the upper bound at 1017 nm. Another isolator is spliced at the output of the pre-amplifier.

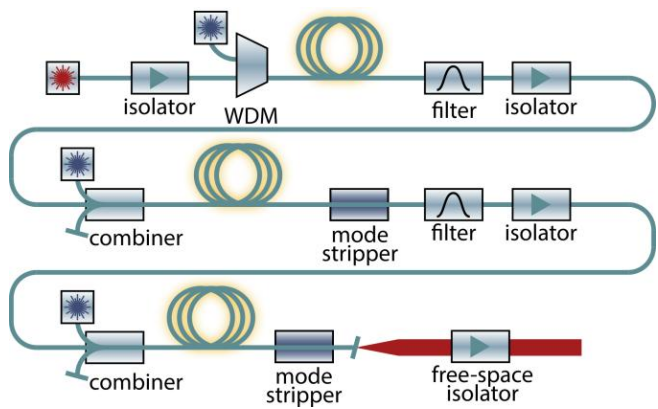

Fig. 1: Tunable laser schematic. All devices are fiber-pigtailed and polarization-maintaining. All pumps are at 976 nm.

A second stage of amplification based on 45 cm of clad-pumped fiber (nLight Liekki Yb 1200 12/125) then ramps up the power to about 2 W before filtering by a tunable, 3.5 nm-wide optical filter to ensure that a clean signal seeds the last stage. A third isolator separates the second and last stage, which is again clad-pumped at 976 nm, and based on 66 cm of highly doped custom silica fiber (MFD~17µm, core diameter ~ 22µm, clad diameter~130, NA=0.075 clad pump absorption 26 dB/m@976 nm, ~2wt% Yb concentration). The small clad/core ratio (about 6) favors emission of short wavelengths [31]. The injected power at the input of the last stage is about 0.5 W at 997 nm, with negligible ASE thanks to the filter at the output of the second stage. For consistency, the seed power of the last stage is kept at 0.5 W over the whole operating range, although it can go as high as 1.5 W for the longest wavelengths (>1010 nm). Furthermore, a higher injection power is not so crucial for this spectral area as the third stage is close to saturation with 500 mW. The second stage remains however of prime significance at these wavelengths as it enables to keep a strong and clean injection signal, which is paramount for the performance of the third stage. The second and third stages are clad-pumped at 976 nm with grating-stabilized diode lasers. We have also explored both clad pumping at 915 nm and 940 nm. However, the figures of merit (mainly efficiency and optical-to-noise ratio (OSNR)) of the amplifier where lower compared to 976 nm pumping. The active fibers are very short in order to limit gain around 1030 nm which would cause parasitic lasing. The length of the active fiber is determined by a cut-back procedure with a tradeoff between efficiency and OSNR. The design criteria for the last stage was an OSNR of 45 dB@1000 nm. In Fig. 2: Efficiency and OSNR vs fiber length@1000 nm.Fig. 2 the efficiency and the OSNR are plotted versus fiber length at a wavelength of 1000 nm. Although Stimulated Brillouin Scattering (SBS) is not an issue with such short lengths, the residual pump powers of clad-pumped stages are rather high. Consequently, the mode-strippers need to be robust and well-designed to handle the large power.

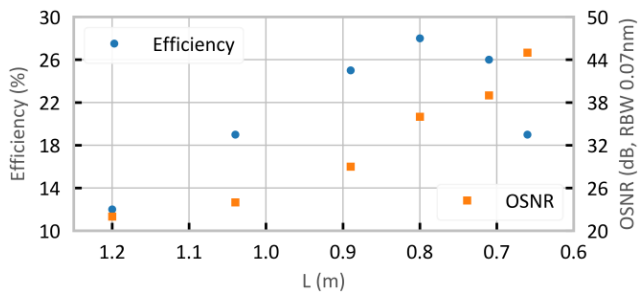

Fig. 2: Efficiency and OSNR vs fiber length@1000 nm.

## 3. Results

The output power vs launched pump power is plotted in Fig. 3. The highest power of 16.5 W is reached for both 1015 nm and 1017 nm, with a slope efficiency exceeding 40 %. Upon tuning the signal wavelength down to 997 nm, the efficiency decreases gradually to 14 %, while still allowing to reach more than 3 W of output power before the onset of parasitic lasing around 1030 nm. The polarization extinction ratio (PER) was measured to more than 17 dB.

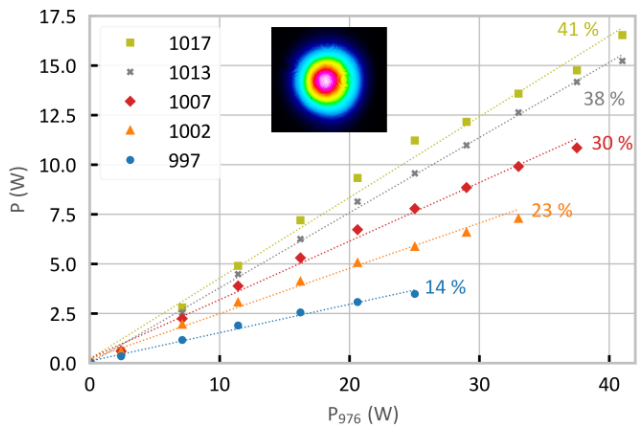

Fig. 3: Output signal power (excluding ASE power) vs pump power for different wavelengths. Slope efficiencies in % are also indicated. Inset: mode at maximum pump power.

The spectra of the output at different operating wavelengths are plotted on Fig. 4. In the worst case (997 nm), the OSNR is as high as 40 dB, which is a remarkable result given that there is no previous work with Yb doped silica fiber amplifiers at this wavelength. Such a high OSNR corresponds to a total signal-to-integrated ASE ratio of 18 dB, i.e. a total ASE power of only 50 mW for the maximum power of 3.5 W (less than 2 %). When tuning the wavelength up, the OSNR keeps increasing until more than 60 dB at 1017 nm (35 dB of signal-to-integrated ASE ratio), which corresponds to a total ASE power of 5 mW for the maximum power of 16.5 W. These figures demonstrate the high spectral quality of the MOPA, even when operating far away from the maximum Yb gain. It is worth remembering that the amplification process even in high power MOPA has nearly no effect on the linewidth of a narrow SF source at the Hz level [8]. This subject has been extensively studied in both the gravitation wave and frequency metrology communities [32]–[34]. Thus, no further investigation is required to characterize the single frequency behavior of the output signal. The output signal then retains the spectral properties of the SF Toptica seed.

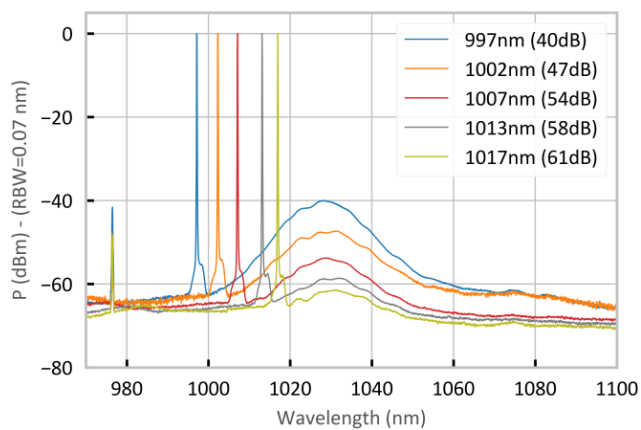

Fig. 4: Spectra for different signal wavelengths the. weak residual 976 nm pump is due to the short active fiber. OSNR are highlighted in the legend.

The efficiencies and OSNR of the system over its whole tuning range are summed up in Fig. 5. Although the MOPA's performance is best when operating closer to the peak gain of Yb, the figures at the lower edge of its operating range are still more than respectable. A key strategy for achieving such respectable OSNR is the cascade of short fiber length amplifiers operated at moderate gain levels. This approach reduces the ASE content and optimize amplifier saturation at the expense of efficiency. In the most critical part of the spectrum i.e. <1005 nm the first (second) stage gain is about 5 (7) dB.

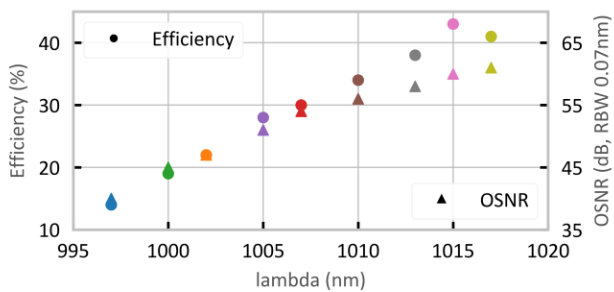

Fig. 5: Efficiency and OSNR for different signal wavelengths.

The efficiency starts to saturate at 1015 nm, indicating that the fiber is just the right length: short enough to allow operation at 997 nm, but long enough not to significantly impair efficiency at longer wavelengths.

Although the OSNR at the lower end of the operating range would strongly benefit from a shorter fiber, this would reduce the performance at the longer wavelengths as well as increase the residual pump power to be disposed of, putting even more pressure on the design of the mode stripper.

We have investigated the performance of several commercial fibers with poorer results in term of efficiency and OSNR.

For robust operation compatible with industrial-grade laser, an OSNR over 40dB is required in order to avoid parasitic lasing or even catastrophic Q-switch events.

The beam quality of our system has been evaluated by a $M^2$ measurements at full power at 1013 nm (Fig. 6). Additional measurements show little to no change with power or wavelength. This gives a good insight about the ability of the amplifier to deliver a good quality beam.

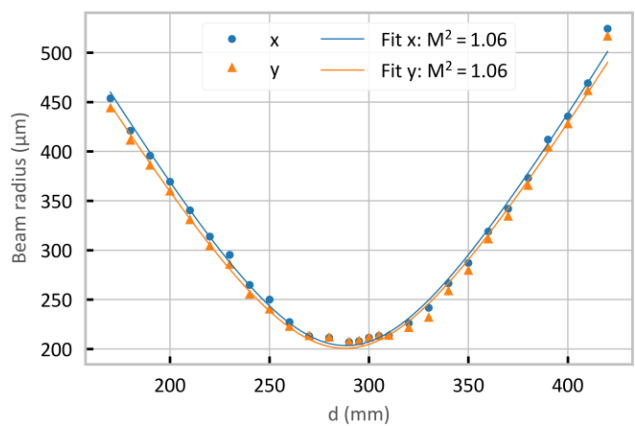

Fig. 6: Beam quality measurement ($M^2$) of the laser beam at 1013 nm at full output power.

## 4. RIN and power stability

The relative intensity noise (RIN) has been characterized over the whole tuning range of the amplifier. The results are shown in Fig. 7. Overall, the performance is excellent; the behavior matches the theoretical model where pump noise dominates at low frequency, while seed noise impacts high frequencies

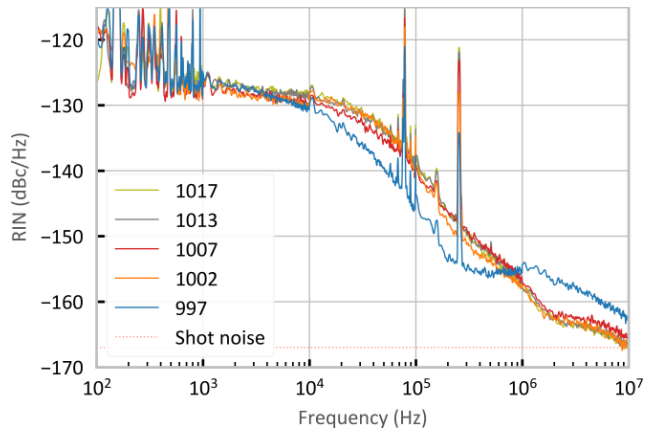

Fig. 7: RIN of output signal at 33 W pump-power for different wavelengths (26 W for 997 nm) (detected power >15 mW). Spikes around 100kHz are electronic pick-up noise.

The corner frequency between the two regimes depends mostly on output power and fiber properties [35], [36].

The slight RIN spectra variations are explained by the increase of the maximum output power with output wavelength, leading to a linear increase of the corner frequency. High-frequency noise can also be impacted by SBS [37].
Here, no degradation is visible as the RIN rolls down to close the shot-noise level of the detected light (20mW) for high frequencies, ruling out the presence of SBS. In addition, no degradation was observed in our previous 50 W@1013 nm amplifier with an even longer active fiber [38].
At 997 nm however, the source laser starts to exhibit some instabilities which impact the output RIN around 1 MHz. However, this slight excess noise does not reflect any inherent issue about the amplification process at this wavelength.
Finally, a long-term stability measurement has been performed, where the laser has been let to operate at 1007 nm at a nominal power of 10 W over 3 days. The trace of the output power is shown in Fig. 8. After a slow power drop due to photodarkening (PD) [39], [40], the power stabilizes just below 8 W. Peak-to-peak stability in this regime is below 5 %, and RMS stability around 1 %. The remaining slow power drifts are correlated to thermal fluctuations in the room, which impair both the laser performance and the power meter response. Eventually, these slow variations can be eliminated with a constant-power feedback circuit.

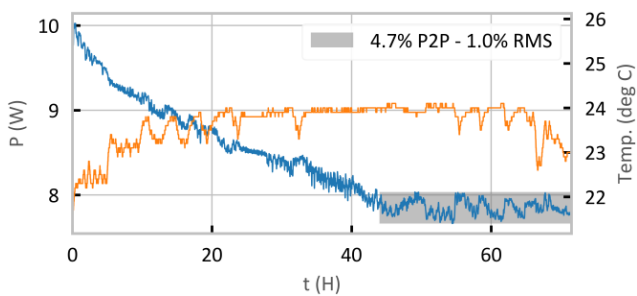

Fig. 8: Long term operation at 1007 nm 10W output power (the first 45-hours are dominated by photodarkening).

Because of the very strong doping concentration and high inversion, the total loss of power after PD is about 22 %. This is significantly higher than 5% power loss measured at 1064 nm operation and can be interpreted as a strong dependency on the inversion level. This is confirmed by the way PD impairs efficiency differently over the spectral operating range; if 22 % of power is lost at 1007 nm, the output decreases only by 15 % at 1017 nm and as much as 29 % at 997 nm. This is because PD has a stronger effect at shorter wavelengths [39]. However, the spectra after PD are barely affected (data not shown), mostly because of the already very weak ASE content in the output signal. We could recover the lost power for the shortest wavelengths as more pump was available (see Fig. 2). Another pump would be needed to retrieve the nominal power at longer wavelengths (>1005 nm).

## 5. Conclusions

In summary, we have developed a robust, high-power, all-fiber, linearly polarized, SF, low-noise MOPA tunable from 997 nm to 1017 nm. Although the laser efficiency eventually suffers from PD, the overall quality of the output (RIN, OSNR, mode) remains unaffected. This novel fiber laser is an excellent replacement for its cumbersome solid-state counterparts, or to replace high-power semiconductor diodes when excellent modal quality is required. Because of the short fiber used, the MOPA would most probably perform equally well in pulsed regime. Furthermore, by delivering these wavelengths in SF regime with a significant amount of power into fiber, a whole range of previously inaccessible wavelength opens up simply by frequency doubling or quadrupling to 500 nm or 250 nm. In addition, by sum frequency generation with an Erbium counterpart, new orange light sources can be generated (~607 nm) with application in quantum optics and quantum memories [41] and by successive frequency doubling in the 304nm region for ozone spectroscopy [42].

## Acknowledgments

This work was supported by Agence Nationale de la Recherche (ANR) (ANR14 LAB05 0002 01), Conseil Régional d'Aquitaine (2017-1R50302-00013493), LAPHIA (Lasers and Photonics in Aquitaine). We acknowledge the technical support of J. H. Codarbox and Ph. Teulat.